\documentclass[aps,prl,twocolumn,superscriptaddress]{revtex4-2}

\usepackage{amsmath,amssymb,amsfonts}
\usepackage{graphicx}
\usepackage{hyperref}

\begin{document}

\title{Quantum typicality survives non-Abelian gauge constraints:\\
exact analytical prediction confirmed in $SU(2)$ lattice gauge theory}


\author{Zhi-Wei Wang}
\email{zhiweiwang.phy@gmail.com}
\affiliation{College of Physics, Jilin University,
Changchun, 130012, People's Republic of China}
\author{Samuel L.\ Braunstein}
\email{sam.braunstein@york.ac.uk}
\affiliation{Computer Science, University of York, York YO10 5GH,
United Kingdom}

\begin{abstract}
Arguments for emergent spacetime require that quantum typicality, the
generic absence of inter-subsystem correlations, persists on the
physical Hilbert space of a gauge theory, where non-Abelian constraints
could in principle inject geometry-supporting entanglement.  Using
$SU(2)$ lattice gauge theory on two-dimensional tori
($d_{\mathrm{phys}}$ up to $4{,}193$), we show that it does: the typical
mutual information between strictly disjoint links matches an exact
parameter-free analytical prediction combining a microcanonical baseline
with Haar-random fluctuations.  The Kogut-Susskind Hamiltonian generates
correlations from states of definite geometry (such as the electric
vacuum), while generic states show only regression to the mean,
establishing that the arrow of correlation growth requires a non-generic
initial condition.
\end{abstract}

\maketitle

\section{Introduction}
\label{sec:intro}

Concentration-of-measure results establish that generic pure states in
high-dimensional Hilbert spaces have vanishing quantum mutual
information between moderate-sized
subsystems~\cite{Lubkin1978,Page1993,Hayden2006,Popescu2006}. Combined
with the conjecture that quantum entanglement is necessary for connected
spacetime geometry~\cite{vanRaamsdonk2010}, this implies that
geometry-supporting states occupy an exponentially thin sliver of the
kinematic Hilbert
space~\cite{WangBraunstein2026PRL,WangBraunstein2026TypeII}.

A central objection is that the physical Hilbert space
$\mathcal{H}_{\mathrm{phys}}$ of a gauge theory or gravitational
system is not the full kinematic space but a constrained subspace
defined by the Gauss law or the Wheeler-DeWitt equation.  If the
constraints preferentially select highly correlated states,
typicality could fail on $\mathcal{H}_{\mathrm{phys}}$.

For non-factorisable Hilbert spaces, typicality bounds have
been shown to persist analytically~\cite{WangBraunstein2026PRL},
with the mutual information decomposing exactly into a
microcanonical baseline and a Dirichlet fluctuation.  However,
existing tests are purely static and do not address the
non-Abelian constraints relevant for gravity, where non-commuting
generators could correlate subsystems during the projection onto
the physical subspace.  Whether the analytical decomposition
survives in this environment, and whether the physical Hamiltonian
can dynamically generate correlations from an initially uncorrelated
state, are both open questions.

\def\leaveout{
For non-factorisable Hilbert spaces (where the subsystem
decomposition involves direct sums rather than tensor products),
typicality bounds have been shown to persist
analytically~\cite{WangBraunstein2026PRL}, with the diagonal
mutual information decomposing exactly into a microcanonical
baseline and a Dirichlet fluctuation.  However, the constraints
relevant for gravity are non-Abelian: the spatial diffeomorphism
and Hamiltonian constraints of general relativity, or the
$SU(2)$ Gauss law in loop quantum gravity, involve
non-commuting generators that could in principle correlate
subsystems during the projection onto the physical subspace.
A key question is whether the analytical decomposition survives
in this non-trivial environment.

A further limitation of existing tests is that they are purely
static: they sample Haar-random states in
$\mathcal{H}_{\mathrm{phys}}$ and check the distribution of
mutual information, but do not address whether the physical
Hamiltonian can dynamically generate correlations from an
initially uncorrelated state.  In the context of emergent
spacetime, the relevant question is not only whether typical
states lack geometry, but whether a physical time evolution can
drive the system from a pre-geometric (uncorrelated) state to a
geometric (correlated) one.}

In this paper we address both the static and dynamic questions using
pure $SU(2)$ lattice gauge theory on two-dimensional periodic lattices.
Working in the spin network basis with the full Kogut-Susskind
Hamiltonian (electric plus magnetic), we demonstrate that typicality
survives the non-Abelian gauge constraints with an exact analytical
match, that the plaquette Hamiltonian generates correlations from the
electric vacuum (and other basis states), and that the arrow of mutual
information growth requires a non-generic (e.g., definite-geometry)
initial condition.

\section{The model}
\label{sec:model}

\subsection{Lattice and link Hilbert space}

We consider an $L_x \times L_y$ periodic lattice (torus) with
$V = L_x L_y$ vertices, $E = 2V$ oriented links (one horizontal
and one vertical per vertex), and $F = V$ plaquettes.

Each link $e$ carries a Hilbert space truncated to
representations $j \leq j_{\max}$ in the Peter-Weyl basis:
\begin{equation}
  \mathcal{H}_e
  = \bigoplus_{j=0}^{j_{\max}}
    \mathbb{C}^{2j+1} \otimes \mathbb{C}^{2j+1},
  \label{eq:link_space}
\end{equation}
with basis states $|j, m_L, m_R\rangle$, where $m_L$ and $m_R$
are the magnetic quantum numbers for the left (source) and right
(target) gauge transformations.  For $j_{\max} = 1/2$, the link
dimension is $d_{\mathrm{link}} = 1^2 + 2^2 = 5$ (one $j = 0$
singlet plus four $j = 1/2$ states).  The kinematic Hilbert space
is $\mathcal{H}_{\mathrm{kin}} = \bigotimes_{e=1}^{E}
\mathcal{H}_e$, with $d_{\mathrm{kin}} = 5^E$.

\subsection{Gauss law}

At each vertex $v$, the Gauss law requires that the total angular
momentum of the adjacent link endpoints vanishes:
\begin{equation}
  \hat{G}_v^a
  = \sum_{e:\,v_s(e)=v} \hat{J}_{e,L}^a
  + \sum_{e:\,v_t(e)=v} \hat{J}_{e,R}^a = 0,
  \label{eq:Gauss}
\end{equation}
for $a = x, y, z$, where $\hat{J}_{e,L}^a$ acts on the left
index of link $e$ and $\hat{J}_{e,R}^a$ on the right index.
The physical subspace
$\mathcal{H}_{\mathrm{phys}} = \bigcap_v \ker(\hat{G}_v^2)$ is
a genuinely correlated subspace, not a tensor product over
vertices.  The correlations arise from the direct-sum structure
of the links: the left and right ends of a link share the same
representation label $j_e$, and the Gauss law forces correlated
superpositions over these shared labels across the
lattice~\cite{WangBraunstein2026PRL}.

\subsection{Spin network basis}

A spin network state is a valid assignment of $j_e \in
\{0, 1/2\}$ to each link such that the Gauss law is satisfied
at every vertex, together with a choice of intertwiner at each
vertex.  The physical subspace dimension $d_{\mathrm{phys}}$ is
computed by exact enumeration of valid spin network
configurations (Table~\ref{tab:dims}).

\begin{table}[ht]
\caption{Physical Hilbert space dimensions for
$j_{\max} = 1/2$ on the torus.}
\label{tab:dims}
\begin{ruledtabular}
\begin{tabular}{ccccc}
Lattice & $V$ & $E$ & $d_{\mathrm{kin}} = 5^E$
& $d_{\mathrm{phys}}$ \\
\hline
$2 \times 2$ & 4 & 8 & $3.9 \times 10^5$ & 63 \\
$2 \times 3$ & 6 & 12 & $2.4 \times 10^8$ & 343 \\
$3 \times 3$ & 9 & 18 & $3.8 \times 10^{12}$ & 4,193 \\
\end{tabular}
\end{ruledtabular}
\end{table}

\subsection{Hamiltonian}

The Kogut-Susskind Hamiltonian~\cite{KogutSusskind1975} consists
of an electric and a magnetic term:
\begin{equation}
  H = g^2 \sum_e \hat{J}_e^2
      - \frac{1}{g^2} \sum_\square
        \mathrm{Re}\,\mathrm{Tr}\,U_\square,
  \label{eq:H}
\end{equation}
where $\hat{J}_e^2 = j_e(j_e+1)$ is the Casimir of the
representation on link $e$, and $U_\square = U_{e_1} U_{e_2}
U_{e_3}^\dagger U_{e_4}^\dagger$ is the Wilson loop around
plaquette $\square$.

The electric Hamiltonian $H_E = g^2 \sum_e j_e(j_e+1)$ is
diagonal in the spin network basis.
The magnetic Hamiltonian $H_B$ is off-diagonal: the plaquette
operator $\mathrm{Tr}\,U_\square$ flips the four links of a
plaquette between $j = 0$ and $j = 1/2$.  Its matrix elements
between spin network states are computed via local tensor
network contractions over the four corners of each
plaquette~\cite{KogutSusskind1975}.

The electric vacuum $|\mathrm{vac}\rangle = |j = 0, \ldots, 0\rangle$ is
the unique state with all links in the trivial representation.  It is a
product state with exactly zero mutual information between any pair of
links, the lattice analogue of a pre-geometric state. Importantly,
because our observable measures the classical MI of the deterministic
$j$-labels, \emph{every} spin network basis state has strictly zero
classical MI initially. (Furthermore, because spin network states are
tensor products of vertex intertwiners, their true quantum MI between
disjoint links is also exactly zero).

\subsection{Mutual information in the spin network basis}

In the spin network basis, the observable associated with a
single link is its representation label $j_e \in \{0, 1/2\}$,
a two-dimensional observable.  The reduced density matrix on a
subsystem $A$ (a set of links) is obtained by tracing over all
links and intertwiners outside $A$.

For strictly disjoint link pairs (links sharing no vertex), the reduced
density matrix $\rho_{AB}$ must commute with the local $SU(2)$ gauge
transformations at each endpoint.  By Schur's Lemma, $\rho_{AB}$ is
maximally mixed within each $(j_A, j_B)$ block, so the full quantum
mutual information reduces exactly to the classical Shannon mutual
information of the $j$-labels. It is mathematically crucial to note that
this exact cancellation of intra-sector quantum entanglement only holds
for strictly disjoint links. For adjacent links sharing a vertex, local
gauge invariance enforced by the intertwiner forces the shared endpoints
into highly entangled pure states (e.g., singlets), heavily suppressing
the joint quantum entropy and leaving a massive quantum entanglement
contribution. Therefore, to cleanly test typicality via the
$j$-labels, one must strictly restrict the analysis to disjoint pairs.

The quantum mutual information between subsystems $A$ and $B$ is
$I(A{:}B) = S(\rho_A) + S(\rho_B) - S(\rho_{AB})$, where $S(\rho) =
-\mathrm{Tr}(\rho \log_2 \rho)$ is the von Neumann entropy.

\section{Results}
\label{sec:results}

We set $g^2 = 1$ throughout.  All mutual information values are
in bits.

\subsection{Static typicality and the analytical prediction}

For each lattice size, we sample 50 Haar-random states in
$\mathcal{H}_{\mathrm{phys}}$ and compute the single-link mutual
information for a strictly disjoint link pair (e.g., parallel horizontal
links separated by at least one lattice spacing), for which the
$j$-label MI equals the full quantum MI by Schur's Lemma on all lattice
sizes.

The typical mutual information admits an exact analytical prediction.
The fully mixed state $\rho_{\mathrm{mixed}} =
\mathbb{I}/d_{\mathrm{phys}}$ on the physical subspace is not
uncorrelated: the Gauss law forces $j = 1/2$ links to form closed flux
loops, injecting a microcanonical baseline of structural correlations.
Let $d_{ab}$ be the number of spin network states with
$j_A = a$ and $j_B = b$ for the chosen disjoint pair.
The microcanonical mutual information is the Shannon MI of the
probabilities $p_{ab} = d_{ab}/d_{\mathrm{phys}}$:
\begin{equation}
  I_{\mathrm{mixed}}
  = \sum_{a,b} p_{ab} \log_2
    \!\left(\frac{p_{ab}}{p_a\, p_b}\right).
  \label{eq:Imixed}
\end{equation}
For a Haar-random pure state, the squared amplitudes in each
subsector follow a Dirichlet distribution, producing a
finite-size fluctuation that, for binary subsystems
($d_A = d_B = 2$), evaluates
exactly~\cite{WangBraunstein2026PRL} to
\begin{equation}
  \langle I \rangle_{\mathrm{fluct}}
  = \frac{(d_A - 1)(d_B - 1)}{2\,d_{\mathrm{phys}}\ln 2}
  = \frac{1}{2\,d_{\mathrm{phys}}\ln 2}.
  \label{eq:Ifluct}
\end{equation}
The total typical MI is the sum of these two contributions:
\begin{equation}
  \langle I(A{:}B)\rangle
  = I_{\mathrm{mixed}}
    + \frac{1}{2\,d_{\mathrm{phys}}\ln 2}.
  \label{eq:Itotal}
\end{equation}

Table~\ref{tab:typicality} shows the comparison.  The
subsector dimensions $d_{ab}$ are obtained by exact enumeration,
and the analytical prediction matches the Monte Carlo data with
no free parameters.

\begin{table}[ht]
\caption{Static typicality: strictly disjoint pair
mutual information for
Haar-random physical states.  $I_{\mathrm{mixed}}$ is the
microcanonical baseline~(\ref{eq:Imixed}),
$I_{\mathrm{fluct}}$ the Dirichlet
fluctuation~(\ref{eq:Ifluct}), and $I_{\mathrm{pred}}$ their
sum.  All values in bits.}
\label{tab:typicality}
\begin{ruledtabular}
\begin{tabular}{cccccc}
Lattice & $d_{\mathrm{phys}}$
& $I_{\mathrm{mixed}}$ & $I_{\mathrm{fluct}}$
& $I_{\mathrm{pred}}$ & $\langle I\rangle_{\mathrm{MC}}$ \\
\hline
$2 \times 2$ & 63 & 0.0036 & 0.0115 & 0.0151 & 0.012 \\
$2 \times 3$ & 343 & 0.0015 & 0.0021 & 0.0036 & 0.004 \\
$3 \times 3$ & 4,193 & 0.0028 & 0.0002 & 0.0030 & 0.003 \\
\end{tabular}
\end{ruledtabular}
\end{table}

The microcanonical baseline $I_{\mathrm{mixed}}$ remains small
($\lesssim 0.004$ bits) across all lattice sizes, while the
Dirichlet fluctuation vanishes as $1/d_{\mathrm{phys}}$.
The typical MI therefore decreases toward a bounded structural
floor reflecting the correlations that the gauge constraints
inject into every physical state.  This floor is a negligible
fraction of the maximum possible MI of 1 bit.

On the $2 \times 2$ and $2 \times 3$ tori, adjacent horizontal
links span a non-contractible cycle, and their MI includes
topological quantum coherences that inflate the observed values
(see Appendix~\ref{sec:topological} for details).
Fig.~\ref{fig:scaling} shows the scaling of both the disjoint
and adjacent pairs, with the analytical
prediction~(\ref{eq:Itotal}) overlaid on the disjoint data.

\begin{figure}[ht]
\includegraphics[width=\columnwidth]{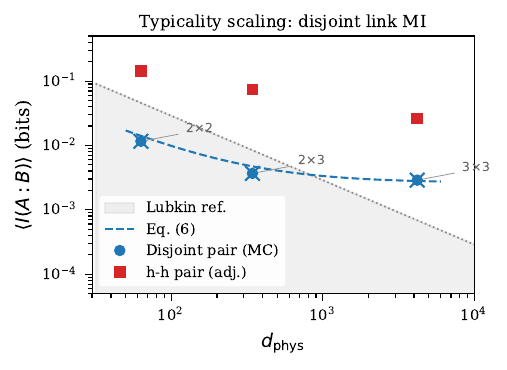}
\vskip -0.15in
\caption{Typical single-link MI as a function of $d_{\mathrm{phys}}$.
Red squares: adjacent h-h pair (includes topological quantum
coherences on $L_x = 2$ lattices). Blue circles: strictly
disjoint pair. Dashed blue curve: parameter-free analytical
prediction~(\ref{eq:Itotal}). Gray shading: unconstrained Lubkin
reference baseline.}
\label{fig:scaling}
\end{figure}

\subsection{Vacuum dynamics}

\begin{figure}[ht]
\includegraphics[width=\columnwidth]{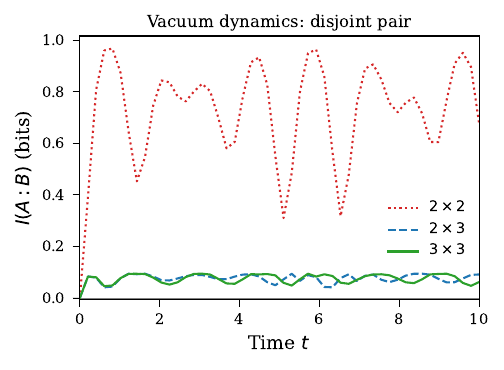}
\vskip -0.15in
\caption{MI between a strictly disjoint link pair as a
function of time, starting from the electric vacuum ($I = 0$).
Red dotted: $2 \times 2$ ($d_{\mathrm{phys}} = 63$);
blue dashed: $2 \times 3$ ($d_{\mathrm{phys}} = 343$);
green solid: $3 \times 3$ ($d_{\mathrm{phys}} = 4{,}193$).
The plaquette Hamiltonian generates inter-link
correlations on a timescale set by the inverse plaquette
energy scale.}
\label{fig:vacuum}
\end{figure}

Starting from the electric vacuum (all links in $j = 0$), we
evolve under the full Hamiltonian~(\ref{eq:H}) and compute the
disjoint pair mutual information as a function of time.
Table~\ref{tab:vacuum} summarises the results and
Fig.~\ref{fig:vacuum} shows the time evolution.

\begin{table}[ht]
\caption{Vacuum dynamics: MI rising from the
pre-geometric vacuum (disjoint pair).  All values in bits.}
\label{tab:vacuum}
\begin{ruledtabular}
\begin{tabular}{cccccc}
Lattice & $d_{\mathrm{phys}}$ & $n_{\mathrm{eig}}$ & $I(0)$
& $I_{\max}$ & $t_{\max}$ \\
\hline
$2 \times 2$ & 63    & 19  & 0 & 0.968 & 0.8 \\
$2 \times 3$ & 343   & 154 & 0 & 0.094 & 8.6 \\
$3 \times 3$ & 4,193 & 920 & 0 & 0.094 & 3.1 \\
\end{tabular}
\end{ruledtabular}
\end{table}

In every case, the MI starts at exactly zero and
rises to a significant fraction of a bit.  The column
$n_{\mathrm{eig}}$ shows the number of distinct eigenvalues of
the full Hamiltonian on $\mathcal{H}_{\mathrm{phys}}$,
confirming that the plaquette term breaks the electric
degeneracies and produces genuinely non-periodic dynamics.

The vacuum couples to multiple eigenstates of $H$ (4 for
$2 \times 2$, 10 for $2 \times 3$), so the time evolution
explores a non-trivial subspace of
$\mathcal{H}_{\mathrm{phys}}$.  The plaquette operator creates
superpositions of $j = 1/2$ flux loops, building up the
inter-link correlations that constitute the lattice analogue of
spatial connectivity.

\subsection{Haar-random drift test}

\begin{figure*}[ht]
\includegraphics[width=\textwidth]{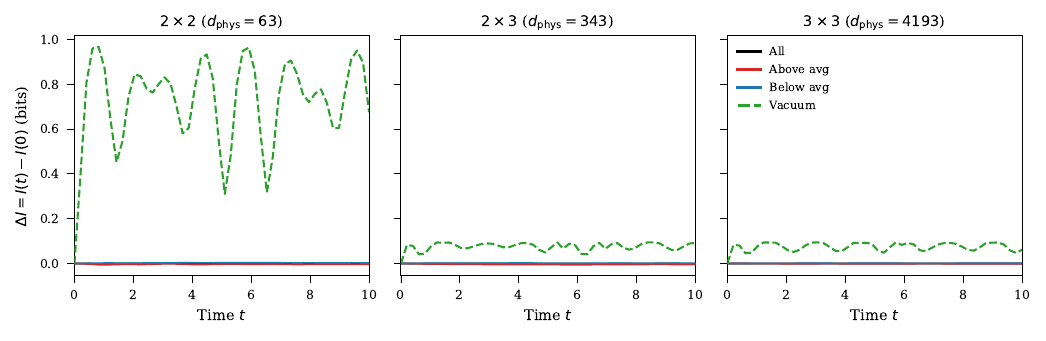}
\vskip -0.15in
\caption{Haar-random drift test (disjoint pair).  Each panel shows the
mean drift $\Delta I = I(t) - I(0)$ for states with
above-average (red) and below-average (blue) initial MI,
the overall mean (black), and the vacuum (green, dashed).
The conditioned drift is symmetric (regression to the mean)
and vanishes with increasing system size.  The vacuum drift
is qualitatively different: it rises from zero,
reflecting genuine correlation generation rather than
statistical regression.}
\label{fig:drift}
\end{figure*}

The vacuum is a special (computational basis) state, meaning its
classical MI is exactly zero. Does the arrow of MI growth persist for
generic highly superposed initial conditions?  To test this, we draw 200
Haar-random initial states on $\mathcal{H}_{\mathrm{phys}}$, evolve each
under $H$, and compute $\Delta I(t) = I(t) - I(0)$ for the disjoint
pair.  We partition the samples into above-average and below-average
initial MI and compute the conditioned drift (Table~\ref{tab:drift} and
Fig.~\ref{fig:drift}).  The mean drift is zero at all lattice sizes,
confirming that the Haar measure is unitarily invariant: there is no
arrow of time from generic initial conditions.  The conditioned drift
shows symmetric regression to the mean: above-average states decrease,
below-average states increase, by approximately equal amounts. This is a
statistical effect, not a dynamical one.

\begin{table}[ht]
\caption{Haar-random drift test: conditioned drift
$\Delta I = I(t) - I(0)$ at $t = 2.0$ for the disjoint pair.
``Above'' and ``Below'' refer to states with initial MI
above or below the Haar mean.  All values in bits.}
\label{tab:drift}
\begin{ruledtabular}
\begin{tabular}{cccccc}
Lattice & $d_{\mathrm{phys}}$ & Mean drift & Above
& Below & Vacuum \\
\hline
$2 \times 2$ & 63 & $+0.000$    & $-0.004$ & $+0.002$ & $+0.844$ \\
$2 \times 3$ & 343 & $+0.000$   & $-0.003$ & $+0.001$ & $+0.069$ \\
$3 \times 3$ & 4,193 & $+0.000$ & $-0.001$ & $+0.001$ & $+0.060$ \\
\end{tabular}
\end{ruledtabular}
\end{table}

The vacuum drift ($+0.060$ at $t = 2$ for $3 \times 3$) is sixty times
the below-average conditioned drift ($+0.001$) and far exceeds the mean
drift (zero). The vacuum is not merely a below-average fluctuation;
however, it is not unique in this regard. Because our
observable measures the classical MI of the $j$-labels, \emph{every}
spin network computational basis state (i.e., any state of definite
geometry or flux) has exactly zero initial classical MI. Starting the
time evolution from any such definite-geometry state will cause the
plaquette operators to immediately create superpositions of flux
configurations, driving the mutual information to grow in the exact same
directed manner. The evolution of these basis states is qualitatively
different from the regression exhibited by generic states.

A search for spontaneous ``nucleation'' of geometry from
generic states (see Appendix~\ref{sec:nucleation}) finds
correlated fluctuations at $d_{\mathrm{phys}} = 63$ but none at
$d_{\mathrm{phys}} = 4{,}193$, consistent with the geometric
sector occupying an exponentially small fraction of
$\mathcal{H}_{\mathrm{phys}}$.

\section{Discussion}
\label{sec:discussion}

\subsection{Typicality survives non-Abelian constraints}

The exact analytical match between Eq.~(\ref{eq:Itotal}) and the Monte
Carlo data (Table~\ref{tab:typicality}) for disjoint links demonstrates
that the decomposition of typical MI into a microcanonical baseline and
a Dirichlet fluctuation, proved for unconstrained spaces in
Ref.~\onlinecite{WangBraunstein2026PRL}, extends to the gauge-invariant
subspace of an $SU(2)$ lattice gauge theory. The gauge constraints
inject a bounded structural correlation ($I_{\mathrm{mixed}} \lesssim
0.004$ bits) that remains bounded, while the Haar-random fluctuation
decays as $1/d_{\mathrm{phys}}$.  This confirms that concentration of
measure applies within $\mathcal{H}_{\mathrm{phys}}$ and supports the
conjecture that typicality survives on the physical Hilbert space of
quantum gravity.

\subsection{Dynamical correlation generation}

The vacuum dynamics (Table~\ref{tab:vacuum}) demonstrates that the
Kogut-Susskind Hamiltonian generates inter-link correlations from the
electric vacuum (and indeed, from any spin network state with definite
$j$-labels) on a timescale set by the inverse plaquette energy scale.
This is the lattice analogue of spacetime emergence: the vacuum is a
pre-geometric state (no flux, no correlations, no connectivity), and the
physical Hamiltonian drives it toward a state with nonzero MI between
spatially separated links.

\subsection{The arrow of mutual information growth}

The Haar-random drift test (Table~\ref{tab:drift}) establishes that the
arrow of MI growth is not a generic feature of the dynamics but requires
a non-generic (e.g., definite-geometry) initial condition. Generic
(Haar-random) states show zero mean drift and symmetric regression to
the mean.  Computational basis states of definite geometry or flux (such
as the vacuum), which are maximally atypical (zero MI, at the extreme
tail of the Haar distribution), show a strong directed increase.

This parallels the situation in classical statistical mechanics, where
the arrow of time requires a low-entropy initial
condition~\cite{Boltzmann1896}.  In our setting, the role of low entropy
is played by low MI (no inter-link correlations, no geometry).  The
typicality argument provides the justification for this initial
condition: if the universe began in a state of definite flux or a
pre-geometric phase (as the concentration-of-measure results suggest),
the natural initial state has zero MI, and the physical Hamiltonian
generates geometry from that starting point. While the arrow of MI
growth is a generic feature of starting from any definite-geometry
state, the vacuum is physically distinguished not by fine-tuning but by
the absence of any structure: in the pre-geometric phase, there is no
manifold to support flux, so the vacuum is the naturally available
state.

The absence of spontaneous nucleation events at
$d_{\mathrm{phys}} = 4{,}193$ is a direct physical
manifestation of the concentration of the Dirichlet
distribution: as the physical Hilbert space dimension grows,
the probability of extreme fluctuations into geometric
configurations is suppressed
exponentially~\cite{WangBraunstein2026PRL}.

\subsection{Limitations}

Several limitations should be noted.  First, the $SU(2)$ Gauss law is
only one of the constraints in quantum gravity; the Hamiltonian
constraint (Wheeler-DeWitt equation) is not gauge-theoretic and could
have different correlation properties. Second, the $j_{\max} = 1/2$
truncation discards higher representations present in the full theory;
however, this truncation is standard in quantum link
models~\cite{Chandrasekharan1997} and preserves exact gauge invariance.
Third, the lattice sizes are small: all tests require exact unitary
dynamics and full ensemble sampling, limiting us to
$d_{\mathrm{phys}} = 4{,}193$ ($3 \times 3$).  The clear scaling trends
across three sizes suggest the results extend to larger systems.

\section{Conclusion}
\label{sec:conclusion}

Using pure $SU(2)$ lattice gauge theory on two-dimensional
periodic lattices, we have shown that non-Abelian gauge
constraints do not rescue typical quantum states from the
absence of geometry.

First, quantum typicality survives the imposition of non-Abelian
gauge constraints: the single-link MI for
Haar-random physical states matches an exact parameter-free
analytical prediction~(\ref{eq:Itotal}) combining a
microcanonical baseline set by the Gauss law with Haar-random
Dirichlet fluctuations, confirming the analytical framework
of the companion papers~\cite{WangBraunstein2026PRL}.

Second, the Kogut-Susskind Hamiltonian (electric plus magnetic)
generates inter-link correlations from the electric vacuum, as well as
from any spin network computational basis state, which all have exactly
zero classical MI.  This provides a concrete mechanism for geometry
emergence from a pre-geometric (or definite-flux) initial condition
within the gauge-invariant sector.

Third, the arrow of MI growth requires a highly atypical unentangled
initial condition: generic (Haar-random) states show no systematic
drift, only regression to the mean. While the arrow of time is a generic
feature of starting from any computational basis state (any state of
definite geometry or flux), the vacuum remains physically distinguished
not by fine-tuning but by the absence of any structure: in the
pre-geometric phase, there is no manifold to support flux, so the vacuum
is the naturally available state.

Fourth, a search for spontaneous nucleation of geometry
from generic states reveals sharp finite-size scaling:
correlated fluctuations occur at $d_{\mathrm{phys}} = 63$ but
are entirely absent at $d_{\mathrm{phys}} = 4{,}193$,
consistent with the geometric sector occupying an exponentially
small fraction of the physical Hilbert space.

These results close a significant gap in the typicality
programme for emergent spacetime: the non-Abelian gauge
constraints characteristic of gravity do not invalidate the
argument, and the physical Hamiltonian provides the mechanism
for geometry to emerge from an initially structureless state.

\vskip 0.1in

\begin{acknowledgments}
\textit{Data availability.}---The spin network Hamiltonian matrices were
generated using custom Python code based on the tensor network
contraction method described in the text.  The simulation scripts for
the typicality sampling, vacuum dynamics, and drift tests are available
from the authors upon request.  No external experimental datasets were
used.
\end{acknowledgments}

\appendix

\section{Classical and quantum contributions to
the mutual information}
\label{sec:classical_quantum}

The mutual information between two links in the spin network
basis has two distinct sources of correlation: inter-sector
coherences (off-diagonal elements of $\rho_{AB}$ between
different $(j_A, j_B)$ blocks) and intra-sector entanglement
(quantum correlations within a single $(j_A, j_B)$ block,
arising from the magnetic quantum numbers).

Inter-sector coherences survive the partial trace only when
the subsystem $A \cup B$ spans a complete non-contractible
cycle on the lattice.  In such cases, flipping the $j$-labels
of all links in the cycle preserves the Gauss law at every
vertex without altering the environment, allowing off-diagonal
terms between different $(j_A, j_B)$ blocks to survive.  When
$A \cup B$ does not span a cycle, flipping the links
necessarily alters the environment, and the partial trace kills
these inter-sector coherences.

Intra-sector entanglement, by contrast, depends on whether
the two links share a vertex.  For strictly disjoint links
(sharing no vertex), the reduced density matrix $\rho_{AB}$
must commute with local $SU(2)$ gauge transformations at each
endpoint independently.  By Schur's Lemma, $\rho_{AB}$ is
maximally mixed within each $(j_A, j_B)$ block, the
intra-sector entanglement vanishes, and the full quantum MI
reduces exactly to the classical Shannon MI of the $j$-labels.
For adjacent links sharing a vertex, the intertwiner at the
shared vertex forces the shared endpoints into highly entangled
states (e.g., singlets), suppressing the joint quantum entropy
within each block and leaving a substantial quantum
entanglement contribution.  In this case, the full quantum MI
exceeds the classical $j$-label MI.

The main text uses strictly disjoint link pairs throughout,
ensuring that the $j$-label MI equals the full quantum MI by
Schur's Lemma.

\section{Topological transition and classical baseline}
\label{sec:topological}

Comparing the h-h and h-v pairs reveals a topological
transition.  On the $2 \times 2$ and $2 \times 3$ tori
($L_x = 2$), the two adjacent horizontal links $h(0,0)$ and
$h(1,0)$ together form a complete non-contractible cycle
around the $x$-direction.  As discussed in Appendix~\ref{sec:classical_quantum},
the off-diagonal coherence between the flux-loop
configurations survives the partial trace, so the h-h MI
on these smaller lattices includes both inter-sector and
intra-sector quantum contributions.  On the $3 \times 3$
lattice ($L_x = 3$), the two links no longer close a cycle,
killing the inter-sector coherences, though intra-sector
entanglement from the shared vertex persists.  The sharp
drop in the h-h MI from the $2 \times 2$ to the
$3 \times 3$ lattice is largely an artefact of this
topological transition.

The main text uses a strictly disjoint pair (parallel
horizontal links sharing no vertex), for which Schur's Lemma
guarantees that the $j$-label MI equals the full quantum MI.
The disjoint pair exhibits values (0.012, 0.004, 0.003) that
decrease monotonically.  This scaling is quantitatively
explained by the analytical prediction Eq.~(\ref{eq:Itotal}):
the microcanonical baseline $I_{\mathrm{mixed}}$ remains
bounded at $\lesssim 0.004$ bits, while the Dirichlet
fluctuation decays as $1/d_{\mathrm{phys}}$.

The subsector dimensions entering the microcanonical baseline are:

\begin{table}[ht]
\caption{Subsector dimensions $d_{ab}$ for the disjoint pair,
where $a = j_{h(0,0)}$ and $b = j_{h(0,1)}$ for two parallel
horizontal links sharing no vertex.}
\label{tab:dab}
\begin{ruledtabular}
\begin{tabular}{cccccc}
Lattice & $d_{\mathrm{phys}}$ & $d_{00}$ & $d_{0,1/2}$
& $d_{1/2,0}$ & $d_{1/2,1/2}$ \\
\hline
$2 \times 2$ & 63    & 8   & 13  & 13  & 29 \\
$2 \times 3$ & 343   & 42  & 73  & 73  & 155 \\
$3 \times 3$ & 4,193 & 510 & 867 & 867 & 1,949 \\
\end{tabular}
\end{ruledtabular}
\end{table}

These dimensions are obtained by exact enumeration of the spin
network basis.  As the lattice grows, the ratios $d_{ab}/d_{\mathrm{phys}}$
converge, and $I_{\mathrm{mixed}}$ approaches a finite asymptote
determined by the local vertex constraint structure.

\section{Nucleation search}
\label{sec:nucleation}

We searched for transient ``nucleation'' events over a time
window $t \in [0, 50]$ (200 time steps, 200 Haar-random
samples): moments at which a Haar-random state spontaneously
fluctuates into a configuration with high mutual information
across multiple link pairs simultaneously.

On the $2 \times 2$ lattice ($d_{\mathrm{phys}} = 63$), 4.5\%
of samples exhibit correlated spikes exceeding 1.5 times the
Haar mean on all four monitored pairs, with two samples (1\%)
reaching twice the mean on all pairs and the most geometric
configuration achieving 9.4 times the mean on a single pair.

On the $2 \times 3$ lattice ($d_{\mathrm{phys}} = 343$), 10.5\%
of samples show correlated 1.5-times spikes, but none reach
twice the mean on all pairs.

On the $3 \times 3$ lattice ($d_{\mathrm{phys}} = 4{,}193$), no
correlated spikes are observed at any threshold.

The rapid suppression with system size is consistent with the
geometric sector occupying an exponentially small fraction of
$\mathcal{H}_{\mathrm{phys}}$, suggesting that spontaneous
nucleation of geometry from a generic state occurs on timescales
that grow exponentially with $d_{\mathrm{phys}}$.  This
suppression is a direct manifestation of the concentration of
the Dirichlet distribution on the probability simplex: as
$d_{\mathrm{phys}}$ grows, the distribution concentrates more
tightly around the microcanonical probabilities, suppressing the
extreme tails that would correspond to geometric configurations.

\end{document}